\documentclass[pra]{revtex4-2}
\usepackage{braket}
\usepackage{graphicx}
\usepackage{color}

\begin{document}

\title{Dependence of measurement outcomes on the dynamics of quantum coherent interactions between the system and the meter}

\author{Tomonori Matsushita and Holger F. Hofmann}
\email{hofmann@hiroshima-u.ac.jp}
\affiliation{Graduate School of Advanced Science and Engineering, Hiroshima University, Kagamiyama 1-3-1, Higashi Hiroshima 739-8530, Japan}

\begin{abstract}
Information about the internal properties of a system can only be obtained through interactions of the system with an external meter. However, such interactions generally result in entanglement between the system and the meter, making it difficult to trace the measurement result back to a specific value of the physical property in the system. It is therefore possible that the outcomes of quantum measurements depend in a non-trivial manner on the dynamics of the measurement interaction, possibly providing a physical explanation for the role of measurement contexts in quantum mechanics. Here, we show that the effects of the measurement interaction on the meter can be described entirely in terms of the quantum coherent system dynamics associated with the back-action on the system. For sufficiently small back-action uncertainties, the physical property of the system is described by a weak value obtained from the Hamilton-Jacobi equation of the back-action dynamics. At higher measurement resolutions, the observed values are determined by quantum interferences between different amounts of back-action. Eigenvalues emerge when the quantum interferences between different back-actions correspond to a Fourier transform in the back-action parameter. We conclude that the values of physical properties obtained in quantum measurements originate from the quantum coherent properties of the back-action dynamics generated by that physical property during an interaction. Measurement outcomes represent elements of the dynamics and cannot be explained by measurement independent elements of reality. 
\end{abstract}

\maketitle

\section{Introduction}
Physics is based on experimental observations that can be obtained in the environment external to the system. These experimental observations are the only available evidence for the internal properties of a system, including its motion in space and time. Before the advent of quantum mechanics, it was assumed that physical properties of the system are independent of their measurements, resulting in the implicit assignment of ``element of reality'' to the outcomes of potential measurements, whether they are actually performed or not \cite{Ein35}. If this viewpoint was correct, it should be possible to explain the relation between different measurements in terms of joint probability distributions for the outcomes observed in separate measurements. However, it has been shown long ago that the Hilbert space formalism is not consistent with such measurement independent assignments of values \cite{Bel64, Koc68, Asp82}. Recent research suggest that the problem might involve the representation of quantum measurements within the quantum formalism \cite{Fra18, Bru18, Pro19}. In particular, the unavoidable decoherence associated with the performance of actual measurements seems to play a crucial role in defining the relation between the outcomes of different measurements \cite{Gue21}. Since all measurements require an interaction between the system and an external meter, it should be possible to explain the apparent contradictions between the outcomes of different measurements by investigating how the quantum formalism describes the effects of an interaction between the system and an external meter. However, this analysis of interactions is complicated by the unclear role of quantum effects in the external meter system. Heisenberg's initial discussion of the measurement interaction focused on the consistency of the uncertainty principle and applied only a very rudimentary quantization of external degrees of freedom that omitted any descriptions of quantum coherences in the meter system \cite{Hei27}. A more general description of the measurement interaction was formulated by Ozawa, resulting in new expressions for measurement uncertainties and the relation between resolution and disturbance \cite{Oza03}. Although this theory is mathematically consistent, there has been some controversy regarding its physical meaning because it is difficult to relate the error values defined by the operator formalism to the experimentally observable statistics defined by quantum state \cite{Wat11, Bus13, DreN14, Bus14, Roz15, Mao19}. As a result, various alternative definitions of resolution and back-action have been proposed to clarify how measurement uncertainties appear in the actual statistics of the measurement outcomes \cite{Mac06, Bus07, Bcm14, Rod18, Rod19, Pat19}.

Soon after Ozawa's formulation of measurement uncertainties, it was pointed out by Hall that the lowest measurement errors were obtained when the measurement results were equal to the weak values of the physical property for each measurement outcome \cite{Hal04}. Hall's observation strongly suggests that weak values represent the best possible estimate of the value of an observable based on the information provided by the initial state and the final measurement outcome, establishing an important link between measurement uncertainties and the status of weak values as accurate representations of physical properties between state preparation and the post-selection of a final state \cite{Aha05, Vai17, Coh18, Mat19}. Ozawa-Hall uncertainties describe the fluctuations of weak values for a specific set of initial and final conditions, identifying the weak values as averages with a specific statistical variance. It has recently been shown that these fluctuations can be observed directly as decoherence effects in a probe qubit weakly interacting with the target observable \cite{Hof21}. Weak values and Ozawa-Hall uncertainties therefore provide a complete description of the statistics of weak interactions. In the case of vanishing Ozawa Uncertainties, weak values thus apply to each individual system, describing a fully reversible unitary transformation of the probe in a weak interaction \cite{Lem22, Dzi19}.  

Although weak values provide an accurate description of sufficiently weak interactions between a system and an external probe or meter, it is not clear why the weak values can differ greatly from the eigenvalues observed in more precise measurements based on much stronger interactions. It is often assumed that these eigenvalues represent measurement independent values of a physical property, motivating a search for mechanisms that might distort the measurement outcomes in a weak measurement. In particular, it has been argued that anomalous weak values might be caused by the non-vanishing disturbance of the system in the measurement interaction \cite{Han21, Ips22}. However, it should be noted that the disturbance is actually much stronger and much harder to control in a measurement that resolves the eigenvalues of a physical property. If weak values originate from the disturbance of the system in a weak measurement interaction, we need to consider the possibility that eigenvalues originate from the disturbance of the much stronger interaction required to observe them. To understand whether this is the case or not, it is necessary to develop a more general theory of measurement that applies equally to the weak and the strong interaction limits. In general, a measurement interaction always creates entanglement between the system and the meter, making it difficult to separate the measurement information transferred to the meter from the disturbance of the system \cite{Pat19, Mat21}. The fundamental problem in this discussion is therefore that a detailed analysis of the way in which the disturbance might shape the measurement outcomes is only possible if we can somehow separate the effects acting on the system from the effects acting on the meter. As we have shown previously, such a separation can be achieved by identifying the value of the physical property characterizing the meter with a parameter describing the amount of change in the system and the value of the physical property in the system with a parameter describing the amount of change in the meter \cite{Mat21}. The entanglement generated in the measurement interaction therefore relates the dynamics in the meter with the back-action dynamics in the system.

In the present paper, we investigate the role of the back-action dynamics by representing the transformation of the meter state in terms of the quantum dynamics of the system. As we shall show in the following, the transformation of the meter state can then be traced back to the phases of unitary transformations of the system state that describe the possible back-action effects. We find that these phases can only be defined with precision when information obtained from a subsequent final measurement is included. A complete theoretical description of the effects of the system on the meter must therefore include all information that could be obtained in subsequent measurements of the system after the interaction has occurred. In the present theory, the outcome $\bra{f}$ of the final measurement thus represents all of the information about the system that may become available in the future. We thus obtain a more general description of post-selected measurements covering both weak and strong interaction limits. Importantly, this description does not involve any unnecessary decoherence effects, allowing us to describe the measurement interaction itself as a maximally coherent quantum process. Effectively, the entanglement between the system and the meter can then be expressed entirely in terms of the back-action dynamics of the system. A central result of this analysis is that the effect of the interaction on the meter is described by a superposition of different system dynamics associated with the quantum coherent superposition of different eigenstates of the meter observable in the initial meter state. We can therefore conclude that the changes of the meter state from which the measurement results are obtained are always caused by quantum interferences between different amounts of disturbance of the system, independent of the strength of the measurement interaction. In fact, high measurement resolutions can only be achieved when there is a considerable quantum mechanical uncertainty of the meter property responsible for the disturbance in both the initial meter state and the state that describes the meter readout. High resolution therefore requires quantum interferences between a much wider range of different back-action dynamics, each of which is described by a very different relation between the target observable and any final state post-selected in a subsequent measurement. However, a precise description of the meter response still depends on information that is only available in subsequent measurements of the system. The resolution of eigenvalues does not result in an automatic decoherence of the system state. Instead, eigenvalues will only be obtained in specific meter readouts that realize quantum interferences between the full range of available back-action strengths. Our results thus demonstrate that the results of weak and strong measurements both originate from the same type of quantum interferences between different back-action dynamics. Ironically, a fully meter independent description of the effects on the meter is only possible in the weak limit. Strong measurements require a significant quantum uncertainty in the back-action dynamics of the system to ensure that the measurement result obtained from the meter readout will be consistent with the maximal information about the target observable obtained the initial state and a possible subsequent measurement of the disturbed system. It is therefore essential that the role of the back-action dynamics is explained in a consistent manner for both weak and strong measurements. 

In the limit of small back-action uncertainties, the meter shift can be approximated by the gradient of the phase of the probability amplitude of $\bra{f}$. This gradient is given  by the weak value of the physical property $\hat{A}$ determined by the initial state $\ket{\psi}_S$, the back-action parameter $\varphi_A$, and the final state $\bra{f}$. Weak values therefore describe the meter response when the back-action is known with sufficient precision to linearize the dependence of the amplitude of $\bra{f}$ on the back-action parameter $\varphi_A$. Different values of the back-action parameter result in different values of $\hat{A}$. If classical laws of physics applied, uncertainties in the back-action parameter $\varphi_A$ would merely result in a corresponding uncertainty of the associated values of $\hat{A}$. However, the quantum formalism describes this situation by coherent superpositions of the different back-action parameters $\varphi_A$ determined by meter preparation and readout. In the limit of highly uncertain back-action parameters $\varphi_A$, each meter shift is associated with a Fourier component of the $\varphi_A$-dependence of the amplitude of $\bra{f}$. These Fourier components are given by the eigenvalues of the observable $\hat{A}$. We thus find that eigenvalues are determined by quantum coherences in the disturbance of the system during the measurement interaction. This result indicates that eigenvalues are measurement dependent because they represent the dynamics of the interaction by which they are observed. Neither weak values nor eigenvalues have any measurement independent reality. Instead, both emerge as a non-classical feature of the dynamics of quantum mechanical interactions.

The rest of the paper is organized as follows. In Sec. \ref{sec:backaction}, we analyze the measurement interaction in terms of the back-action representation and show that the effect of the interaction on the meter is described by the back-action dynamics of the system. In Sec. \ref{sec:metershift}, we show that the meter shift is described by weak values obtained from the Hamilton-Jacobi equation of the back-action dynamics for the limit of small back-action uncertainties. In Sec. \ref{sec:uncertainties}, we derive a trade-off relation between the resolution error $\delta A_M$ of the meter and the fluctuations $\delta A_S$ of the physical property $\hat{A}$ associated with the uncertainty of the back-action dynamics in the system. We can then identify the minimal meter error at which the back-action dependent weak value of $\hat{A}$ can be observed. In Sec. \ref{sec:eigenvalues}, we show that the high resolution limit of quantum measurements corresponds to a Fourier transform of the back-action dynamics, where the Fourier component is determined by the meter shift observed in the readout. Eigenvalues thus emerge as Fourier components of the dynamics when the uncertainty of the back-action is very high. In Sec. \ref{sec:SG}, we apply our theory to the Stern-Gerlach experiment and explain the physics of the back-action for this familiar example. In Sec. \ref{sec:finalcondition}, we discuss the relation between uncertainties of the back-action and the final condition $\bra{f}$ in terms of contextuality. Section \ref{sec:conclude} summarizes the results and concludes the paper.

\section{The back-action representation of quantum measurements}
\label{sec:backaction}
A measurement of a target observable $\hat{A}$ in a physical system requires an interaction between the system and an external meter. The value of the target observable $\hat{A}$ can be obtained from the meter response to a force exerted by the system. In the quantum formalism, this force is represented by a unitary transformation that conserves $\hat{A}$ while generating entanglement between the Hilbert space of the system and that of the meter. This entanglement makes it impossible to separate the action of the system on the meter from the back-action of the meter on the system \cite{Mat21}. As a result, the value of the system property $\hat{A}$ that determines the meter response depends in a non-trivial manner on the dynamics of the back-action from the meter on the system. It seems that this aspect of the measurement interaction is not sufficiently understood, especially in the light of arguments that unusual measurement results such as anomalous weak values might actually be caused by the back-action \cite{Han21, Ips22}. In the following, we will therefore represent the effect of the interaction on the meter using the back-action dynamics conditioned by a physical property of the meter. It is then possible to identify the relation between the internal dynamics of the system and the forces exerted by the system on the meter.

An ideal measurement interaction sensitive to only a single system property $\hat{A}$ can be described by a bilinear Hamiltonian where the meter response is generated by a physical property of the meter $\hat{B}$. The Hamiltonian is then proportional to the dyadic product $\hat{A} \otimes \hat{B}$. As discussed in \cite{Mat21}, additional terms in the Hamiltonian would merely add unnecessary errors into the measurement of $\hat{A}$. In general, the Hamiltonian will be time dependent because the measurement must be completed in a finite amount of time. The unitary transformation describing the complete measurement interaction can thus be obtained from the time integral of the interaction Hamiltonian. The interaction of a measurement of the target observable of the system $\hat{A}$ is then given by the unitary operator
\begin{equation}
\label{eq:interaction unitary}
\hat{U}_{SM} = \exp \left( - \frac{i}{\hbar} g \hat{A} \otimes \hat{B} \right),
\end{equation}
where $g$ determines the effective strength of the interaction. Eq.(\ref{eq:interaction unitary}) represents both the action from the system on the meter and the back-action from the meter on the system in a symmetric form in which the operators $\hat{A}$ and $\hat{B}$ are simultaneously generators of the dynamics in one system and measures of the magnitude of the dynamics in the other \cite{Mat21}. Since the purpose of the measurement is to determine the value of $\hat{A}$, the conventional interpretation of the interaction focuses on the dynamics of the meter conditioned by an eigenstate $\ket{a}$ of $\hat{A}$ in the system. However, this approach makes it difficult to identify the back-action dynamics in the system. When the eigenstates of $\hat{A}$ are used in the expansion of the interaction in Eq.(\ref{eq:interaction unitary}), the back-action is hidden in the coherence between the eigenstates associated with the dynamics of the meter. To shift the focus of our analysis to the dynamics in the system, we instead expand Eq.(\ref{eq:interaction unitary}) in terms of eigenstates $\ket{b}$ of the meter observable $\hat{B}$ to express the interaction in terms of the back-action dynamics generated by the target observable $\hat{A}$ in the system. The back-action representation of the measurement interaction is thus given by
\begin{equation}
\label{eq:back-action}
\hat{U}_{SM} = \sum_b \exp \left( - \frac{i}{\hbar} g B_b \hat{A} \right) \otimes \ket{b} \bra{b}.
\end{equation}
The dynamics of the system is described by a superposition of conditional unitary transformations generated by the physical property of the system $\hat{A}$ and determined by eigenvalues $B_b$ of $\hat{B}$ in terms of eigenstates $\ket{b}$ of $\hat{B}$. Eq.(\ref{eq:back-action}) thus allows us to discuss the role of the back-action dynamics in the system in more detail. It should be noted that Eq.(\ref{eq:back-action}) is still a complete description of the measurement interaction between the system and the meter. However, the force exerted on the meter is now expressed in terms of quantum coherence between the system dynamics conditioned by the eigenvalues $B_b$.

We now need to consider the effects of the back-action on the initial state of the system $\ket{\psi}_S$. In general, this initial state can be expressed as a quantum coherent superposition of the eigenstates $\ket{a}$ of $\hat{A}$. Since the goal of our analysis is to provide the most detailed description of the physics of a measurement interaction, we assume that maximal information about the initial conditions of the system is available and the initial state of the system can be described by a pure state. This requires us to interpret possible mixed state inputs as statistical mixtures of a specific set of pure states. Although this is in principle always possible, it should be noted that the same mixed state can be represented by different mixtures of pure states. In general, the missing information of a mixed state should therefore be represented by entanglement with a hypothetical ancilla system. Importantly, all possible decompositions of the mixed state into a mixture of pure states must be consistent with the statistics of effects on the meter system in the measurement. In the following, we will develop the theory for the case of maximal information. However, it should be kept in mind that any lack of information creates an ambiguity that is best represented by quantum entanglement and not by classical statistics.

The back-action will have a non-trivial effect on the quantum coherence between the eigenstate components $\ket{a}$ that characterize the state of the system, and this transformation of the system state will be observed in all future measurements on the system. Since the meter state has to be in a superposition of eigenstates $\ket{b}$, the back-action effects are usually interpreted as a randomization of the phases of the eigenstates $\ket{a}$, resulting in the decoherence associated with a measurement of $\hat{A}$. However, it is possible to get more detailed information about the back-action effects and their quantum coherence from the outcome $\bra{f}$ of a possible future measurement of the system. In the detailed analysis of the measurement interaction, this information about future measurement outcomes is just as important as the information about the precise initial state of the system. If no subsequent measurement on the system is performed, it may be tempting to assume that the correct value of $\hat{A}$ can always be obtained by a subsequent measurement of the unchanged value of $\hat{A}$ after the measurement interaction. In this standard interpretation of quantum measurements, it is assumed that the effects on the meter can be explained by the eigenvalues of $\hat{A}$. However, it should be kept in mind that this interpretation pre-supposes that the only valid subsequent measurement on the system is a precise measurement of $\hat{A}$. The effects on the meter can then be represented by
\begin{equation}
\label{eq:effect on M for a}
\bra{a} \hat{U}_{SM} \ket{\psi}_S = \braket{a|\psi}_S \sum_b \exp \left( - \frac{i}{\hbar} g A_a B_b \right) \ket{b} \bra{b}.
\end{equation}
This operation is a conditional unitary transformation of the meter where the magnitude of the force acting on the meter is expressed by the eigenvalue $A_a$ of $\hat{A}$ determined in the final measurement of $\bra{a}$ in the system. If no other information about future measurements on the system is available, the statistics of the meter response will always be consistent with a corresponding probability distribution over the potential outcomes $\bra{a}$. However, this interpretation is not sufficiently general, because it is not consistent with other subsequent measurements of the system. To understand the general effects of the measurement interaction, it is therefore necessary to consider the information that can be obtained by a completely arbitrary subsequent measurement outcome $\bra{f}$. It should be noted that this measurement outcome is a representation of all of the information about the system that may become available in the environment at a later time, independent of the way in which this may happen. The state vector $\bra{f}$ defines the maximal information that anyone can obtain about the future of the system. If no subsequent measurement is performed, the meter statistics should not be interpreted in terms of a specific subsequent measurement, since the coherence between different outcomes $\bra{a}$ is not lost, but merely unknown. A sufficiently general description of the maximally available information about the measurement interaction therefore requires the assumption of an arbitrary subsequent measurement outcome $\bra{f}$ represented by a coherent superposition of different eigenstates of $\hat{A}$. We thereby take into account the dependence of the interaction effects on information about the coherences between different eigenstates that may be obtained in future interactions of the system with its environment, where $\bra{f}$ provides a compact summary of the information thus obtained. The effects of the measurement interaction of the meter can then be described in terms of coherent amplitudes for each back-action parameter $g B_b$ without selecting a specific eigenstate component $\bra{a}$. This general description of the meter shift associated with an arbitrary final state $\bra{f}$ is given by 
\begin{equation}
\label{eq:effect on M}
\bra{f} \hat{U}_{SM} \ket{\psi}_S = \sum_b \bra{f} \exp \left( - \frac{i}{\hbar} g B_b \hat{A} \right) \ket{\psi}_S \ket{b} \bra{b} .
\end{equation}
The eigenvalues of this operator are given by the evolution of the amplitude of $\bra{f}$ in the initial state of the system $\ket{\psi}_S$ under the conditional unitary transformation generated by $\hat{A}$ and determined by an eigenvalue $B_b$ of $\hat{B}$. The effect of the operator depends on the dependence of its eigenvalues on the eigenvalues $B_b$ of $\hat{B}$. The meter response is thus determined by the dependence of the probability amplitude of $\bra{f}$ on the parameter of the dynamics generated by the target observable $\hat{A}$.

The probability amplitude of $\bra{f}$ in the system represents the evolution of the system state $\ket{\psi}_S$ under the back-action dynamics conditioned by the eigenvalues $B_b$ of $\hat{B}$. It is convenient to summarize the interaction strength $g$ and the eigenvalue $B_b$ in a single continuous parameter $\varphi_A = g B_b$ describing the dependence of the amplitude of $\bra{f}$ on the magnitude of the back-action. It is then apparent that the physical property $\hat{A}$ only appears in the meter responses because the dependence of the magnitude of $\bra{f}$ on $\varphi_A$ involves the generator of the dynamics $\hat{A}$. In the next section, we will take a closer look at the relation between the back-action dynamics generated by $\hat{A}$ and the magnitude of the meter response from which a measurement value of $\hat{A}$ can be determined.

\section{Definition of the meter shift by the back-action dynamics}
\label{sec:metershift}
The effect of the interaction on the meter is described by the probability amplitude between the initial state of the system $\ket{\psi}_S$ and the final state $\bra{f}$ under the conditional unitary transformation determined by the magnitude of the changes in the system state given by the back-action parameter $\varphi_A = g B_b$. The probability amplitude of the system is the eigenvalue of the transformation of the meter state for the eigenstate $\ket{b}$ of $\hat{B}$. The back-action dynamics thus determines the changes of the meter state through its dependence on the parameter $\varphi_A = g B_b$.

The operator in Eq.(\ref{eq:effect on M}) is not necessarily a unitary operator. However, it can be decomposed into a product of a unitary operator and a self-adjoint operator by separating the probability amplitudes into their absolute values and a phase factor,
\begin{equation}
\label{eq:separation}
\bra{f} \exp \left( - \frac{i}{\hbar} g B_b \hat{A} \right) \ket{\psi}_S = \sqrt{P(\psi, f, \varphi_A)} \exp \left( \frac{i}{\hbar} S(\psi, f, \varphi_A) \right).
\end{equation}
Here, $P(\psi, f, \varphi_A)$ is the probability of finding the final state of the system $\bra{f}$ in the initial state $\ket{\psi}_S$ under the unitary transformation determining the magnitude of the changes in the system state $\varphi_A$, and $S(\psi, f, \varphi_A)$ is an action defined by multiplying the phase of the probability amplitude by Dirac's constant $\hbar$. Because the probability $P(\psi, f, \varphi_A)$ depends on the parameter $\varphi_A$, the observation of the final outcome $\bra{f}$ provides information about the value of $\varphi_A = g B_b$.  In the meter system, this information is represented by a self-adjoint operator describing a measurement of the meter observable $\hat{B}$. The self-adjoint operator represents the minimal amount of disturbance of the meter state necessary for the transfer of information about the meter to the system. In terms of the meter dynamics, this disturbance corresponds to a decoherence between the eigenstates $\ket{b}$ of the meter that can be identified with uncontrollable fluctuations in the meter dynamics induced by uncertainties in the target observable $\hat{A}$ \cite{Hof21,Mat21}.

The actual meter shift is described by the unitary operator in Eq.(\ref{eq:separation}). This unitary operation is characterized by the action $S(\psi, f, \varphi_A)$ and reads
\begin{equation}
\label{eq:action on M}
\hat{U}_M = \sum_b \exp \left( \frac{i}{\hbar} S(\psi, f, g B_b) \right) \ket{b} \bra{b}.
\end{equation}
Since the meter shift corresponds to the experimentally observed value of the physical property $\hat{A}$, it should be possible to determine the value of $\hat{A}$ from the form of the action $S(\psi, f, \varphi_A)$. As we saw in Eq.(\ref{eq:effect on M for a}), a well-defined value of $A_a$ is associated with an action of $S = - A_a \varphi_A$. The value of $\hat{A}$ that determines the meter shift can then be derived using the negative derivative of the action in the parameter $\varphi_A$. In the more general case of a final state $\bra{f}$, the dependence of the action $S(\psi, f, \varphi_A)$ on the parameter $\varphi_A$ will be non-linear and the derivative will be a function of $\varphi_A$ itself. However, it is still possible to identify the value of $\hat{A}$ that determines the meter shift with the negative action derivative if the uncertainty of $\hat{B}$ in the meter is sufficiently small so that the action can be linearized within the relevant interval of back-action parameter $\varphi_A$. The meter shift near a specific value of $\varphi_A$ is then given by
\begin{equation}
\label{eq:HamiltonJacobi}
A(\psi, f, \varphi_A) = - \frac{\partial S(\psi, f, \varphi_A)}{\partial \varphi_A}.
\end{equation}
If the quantum uncertainty of the back-action parameter $\varphi_A$ is sufficiently small, the value of $\hat{A}$ that determines the meter response is given by the derivative of the action describing the phase dynamics of the amplitude of $\bra{f}$ generated by $\hat{A}$. This relation between the value of a generator and the derivative of an action functional is equal to the relation between energy and the time derivative of the principal function in the Hamilton-Jacobi equation. Eq.(\ref{eq:HamiltonJacobi}) therefore represents the Hamilton-Jacobi equation of the back-action dynamics, where $\hat{A}$ corresponds to the energy and the back-action parameter $\varphi_A$ corresponds to time. However, this version of the Hamilton-Jacobi equation is based on quantum mechanics and corresponds to the exact phase dynamics of the Schr$\mathrm{\ddot{o}}$dinger equation describing the back-action dynamics. According to Eq.(\ref{eq:separation}), the action is given by 
\begin{equation}
\label{eq:action functional}
S(\psi, f, \varphi_A) = \hbar \mbox{Arg}\left(\bra{f} \hat{U}_{\mathrm{back}}(\varphi_A) \ket{\psi}_S \right),
\end{equation}
where
\begin{equation}
\label{eq:backaction}
\hat{U}_{\mathrm{back}} = \exp\left(-\frac{i}{\hbar} \varphi_A \hat{A} \right)
\end{equation}
is the unitary transformation describing the back-action dynamics. The exact quantum mechanical solution of the Hamilton-Jacobi equation Eq.(\ref{eq:HamiltonJacobi}) can now be obtained from the derivative of this unitary transformation. The result is the real part of the weak value,
\begin{equation}
\label{eq:weakvalue}
- \frac{\partial S(\psi, f, \varphi_A)}{\partial \varphi_A} = \mbox{Re} \left( \frac{\bra{f} \hat{A} \hat{U}_{\mathrm{back}}(\varphi_A) \ket{\psi}_S} {\bra{f} \hat{U}_{\mathrm{back}}(\varphi_A) \ket{\psi}_S} \right).
\end{equation}
It should be noted that only the real part of the weak value is obtained because it describes the unitary part of the meter response. The imaginary part of the same weak value describes the change of the probability $P(\psi,f,\varphi_A)$ that characterizes the weak response of the system to the interaction with the meter \cite{Hof11,Dre12}. Here, we are only concerned with the real part of the weak value, since this part of the weak value can be observed in the magnitude of the meter shift, corresponding to the classical expectation that the physical quantity $\hat{A}$ is the cause of the observable change in the meter.

We would like to emphasize that the meter shift described by the real part of the weak value depends on the back-action parameter $\varphi_A$. Since the analysis is not limited to weak interactions, the back-action parameter $\varphi_A$ can have a significant effect on the outcome $\bra{f}$ of a subsequent measurement. Even in the classical limit, this effect can introduce a dependence of the value of $\hat{A}$ observed for a specific subsequent outcome $\bra{f}$ on the precise back-action induced by the meter observable $\hat{B}$. It may be worth considering the physics of this dependence more carefully. In order to have maximal information about the system, it was necessary to introduce the final state $\bra{f}$. However, this final state is obtained after the back-action dynamics changed the initial state. A fixed selection of $\bra{f}$ therefore represents different final conditions for different back-action parameters $\varphi_A$. The quantum mechanical problem is that a high measurement resolution requires a corresponding uncertainty of the back-action parameter $\varphi_A$ \cite{Mat21}. It is therefore impossible to resolve the back-action dependence of the meter shift in a single measurement. An experimental verification of $A(\psi, f, \varphi_A)$ would require joint measurements of the meter shift and its generator $\hat{B}$. Since a high resolution of $\hat{B}$ is necessary to resolve the back-action dependence of $A(\psi, f, \varphi_A)$, the meter shift will be observed at low resolution and a large number of measurements will be necessary to find the correct value. This method corresponds to a weak measurement, but the weakness of the signal is caused by a specific readout of the meter that combines information about the meter shift with information about the back-action.

The weak values determined from the Hamilton-Jacobi equation describe the dependence of the meter shifts on the back-action parameter $\varphi_A$. This dependence describes a fundamental problem caused by the system-meter entanglement. Any measurement readout that is sufficiently precise to identify the value of $\hat{A}$ changes the future of the system in such a way that the correlation between $\hat{A}$ and a future measurement $\bra{f}$ is now unclear. However, such a relation should be exist if the back-action parameter $\varphi_A$ was known. The weak values thus express a more complete causality relation between the measurement outcomes and the future of the system than a highly resolved measurement can confirm. However, this complete causality relation can only be observed when the resolution of the meter readout is too low to observe the meter shifts in a single measurement. In the following, we will analyze this limitation of quantum measurements in more detail to find out how the back-action dynamics determines meter shifts in the presence of non-vanishing uncertainties in the back-action parameter $\varphi_A$.

\section{Uncertainty trade-off between resolution and back-action}
\label{sec:uncertainties}
The Hamilton-Jacobi equation (\ref{eq:HamiltonJacobi}) determines the value of the meter shift as long as the action is approximately linear in $\varphi_A$ for the relevant range of eigenvalues of $\hat{B}$ in the meter. This range of eigenvalues can be controlled by a combination of meter preparation and readout. It is particularly interesting to consider the possibility of meter readouts that provide simultaneous information on the meter observable $\hat{B}$ and the meter shift. It is then possible to confirm that the meter shift depends on the back-action parameter $\varphi_A$ as described by the Hamilton-Jacobi equation (\ref{eq:HamiltonJacobi}). However, it must be kept in mind that such joint measurements are uncertainty limited. In general, a narrow range of values for $\varphi_A$ necessarily limits the resolution of $A(\psi, f, \varphi_A)$ in the readout of the meter shift.

We can use the dependence of the value $A(\psi, f, \varphi_A)$ on the back-action parameter $\varphi_A$ to identify the maximal resolution at which the observed meter shift can still be identified with a specific value of $A(\psi, f, \varphi_A)$. The rate at which the value of $A(\psi, f, \varphi_A)$ changes when the back-action parameter is varied is given by
\begin{equation}
\label{eq:dependence of meter shift}
\frac{\partial A(\psi, f, \varphi_A)}{\partial \varphi_A} = - \frac{\partial^2 S(\psi, f, \varphi_A)}{\partial \varphi_A^2}.
\end{equation}
As shown in \cite{Mat21}, the uncertainty $\delta \varphi_A$ of the back-action parameter defines a lower bound of the resolution error $\delta A_M$ caused by fluctuations in the meter,
\begin{equation}
\label{eq:meter resolution}
\delta A_M \ge \frac{\hbar}{2 \delta \varphi_A}. 
\end{equation}
This resolution error can now be compared to the intrinsic uncertainty of the value $A(\psi, f, \varphi_A)$ associated with the uncertainty $\delta \varphi_A$ of the back-action parameter. Under the assumption that the fluctuations $\delta \varphi_A$ are sufficiently small, we can estimate this uncertainty using the derivative in Eq.(\ref{eq:dependence of meter shift}),
\begin{equation}
\label{eq:uncertainty of meter shift}
\delta A_S = \left | \frac{\partial^2 S(\psi, f, \varphi_A)}{\partial \varphi_A^2} \right| \delta \varphi_A.
\end{equation}
The uncertainty of the meter shift $\delta A_S$ describes the precision with which the initial condition $\ket{\psi}_S$ and the final condition $\bra{f}$ determine the value of $A(\psi, f, \varphi_A)$ independent of any information obtained in the meter readout. On the other hand, $\delta A_M$ describes the resolution determined by the combination of meter state preparation and readout. The two uncertainties therefore apply to complimentary sources of information about the value of $\hat{A}$. If $\ket{\psi}_S$ and $\bra{f}$ are known, the values of $\hat{A}$ that determine the meter shift can be estimated with an uncertainty of $\delta A_S$, resulting in corresponding fluctuations of the observed values for that combination of $\ket{\psi}_S$ and $\bra{f}$. The meter adds a resolution error of $\delta A_M$ to these fluctuations. The total fluctuation $\delta A$ of the meter readout observed for a specific combination of conditions $\ket{\psi}_S$ and $\bra{f}$ are therefore given by
\begin{equation}
\label{eq:total uncertainty}
(\delta A)^2 = (\delta {A_S})^2 + (\delta {A_M})^2.
\end{equation}
This fluctuation can be observed experimentally in the statistics of the meter readout. The average meter shift is always given by $A(\psi, f, \varphi_A)$, so $\delta A$ determines the uncertainty in the observation of the weak value $A(\psi, f, \varphi_A)$ at an average back-action of $\varphi_A$. The intrinsic uncertainty $\delta A_S$ is proportional to the uncertainty of the back-action $\varphi_A$, making it desirable to reduce the fluctuations of the back-action effect. However, the lower bound of the meter resolution $\delta A_M$ is proportional to inverse of $\delta \varphi_A$, requiring an increases in the fluctuations of the meter readout when the fluctuations of the back-action are reduced. The fluctuations of the meter readout conditioned by the initial condition $\ket{\psi}_S$ and the final condition $\bra{f}$ achieve their minimal value when the contributions of $\delta A_S$ and of $\delta A_M$ are equal. The minimal fluctuation of the readout is then given by
\begin{equation}
\label{eq:control limit}
(\delta A)^2 \ge \hbar \left | \frac{\partial^2 S(\psi, f, \varphi_A)}{\partial \varphi_A^2} \right |. 
\end{equation}
The back-action dynamics associated with the final condition $\bra{f}$ thus determine the lower bound of the meter readout fluctuations around the value $A(\psi, f, \varphi_A)$ for an average back-action of $\varphi_A$. If the meter readout is optimized and the fluctuations of the back-action parameter $\varphi_A$ are determined entirely by the meter preparation, the average back-action will be zero and $\delta A$ describes the fluctuations of the meter readout around the weak value determined by an initial state $\ket{\psi}_S$ and a final post selection of $\bra{f}$. Eq.(\ref{eq:control limit}) thus shows that the resolution at which weak values can be observed depends on the back-action dynamics describing the disturbance with respect to the probability amplitude of the post-selected state. 

In a sufficiently large system, the weak value will correspond to the classically expected values for these initial and final conditions and the fluctuations will often be negligibly small. Weak values can thus bridge the gap between quantum physics and classical physics. A sufficiently simple example may help to illustrate this point. Consider a particle of mass $m$ propagating from $\psi = x_1$ at time $t_1 = 0$ to $f = x_2$ at time $t_2 = t$. At time $t_m = t/2$, the particle interacts with a meter measuring the position $\hat{A} = \hat{x}_m$. The back-action can be expressed by a transfer of momentum, $\varphi_A = p$. The action associated with the amplitudes of $\bra{x_2}$ after a back-action of $p$ is given by
\begin{eqnarray}
S(x_1,x_2,p) &=& \hbar \mbox{Arg}\left(\bra{x_2} \hat{U}_{\mathrm{back}}(p) \ket{x_1}\right)
\nonumber \\
&=& \frac{m}{t} (x_1^2+x_2^2) - \frac{t}{8 m}\left(p+\frac{2m}{t}(x_1+x_2)\right)^2 - \frac{\pi \hbar}{4}.
\end{eqnarray}
The Hamilton-Jacobi equation (\ref{eq:HamiltonJacobi}) determines the value of the position at time $t_m = t/2$ as
\begin{equation}
x_m(x_1,x_2,p) = \frac{1}{2}(x_1+x_2) + \frac{t}{4m} p. 
\end{equation}
This is equal to the classical expectation that the particle moves along a straight line, except for the momentum kick at $t_m = t/2$ which corrects for any deviation of the position $x_m$ from the average value of $(x_1+x_2)/2$. However, this straight line trajectory cannot be observed with arbitrary resolution. As shown in Eq.(\ref{eq:control limit}), the minimal meter fluctuations are given by
\begin{equation}
\label{eq:particle}
\delta x_m \ge \sqrt{\frac{\hbar t}{4 m}}.
\end{equation}
In the macroscopic limit, this relation confirms that we can observe motion in a straight line with only negligibly small deviations owing to the smallness of $\hbar$. However, quantum fluctuations play a significant role in free space particle propagation once we attempt to control the trajectories with a precision higher than the one given in Eq.(\ref{eq:particle}). It might be worth noting that this is the regime in which superpositions of position and momentum can be used to demonstrate that quantum particles do not propagate in straight lines, as shown in a recent experiment \cite{pp1, pp2, Ono22}. The uncertainty limit given in Eq.(\ref{eq:control limit}) may thus be useful in the evaluation of quantum interference effects between the states $\ket{\psi}_S$ and $\bra{f}$ directly observable in projective measurements of $\hat{A}$. In general, $\delta A$ seems to describe the limits of classical causality in the relation between initial and final conditions with the property $\hat{A}$, where weak values provide the natural analog of classical causality relations.

Since weak values appear as quasi-classical descriptions of causality in the present analysis, it may be interesting to consider the relation between the present results and the original theory of weak measurements \cite{Aha88}. In the formulation of that theory, it seems sufficient to analyze the average meter readout, without any regard for the information gained in a single measurement. Thanks to Eq.(\ref{eq:control limit}), we now have a specific uncertainty limit for the observation of weak values in a single meter readout, where the precision achieved depends on the effects of the back-action dynamics on the amplitude of the post-selected state $\bra{f}$. In sufficiently large systems, weak values then appear as the best estimates of the observed values, with a statistical error of $\delta A$ that is usually negligibly small due to the smallness of $\hbar$. Interestingly, the lower bound of the fluctuation $\delta A$ also depends on the action $S(\psi, f, \varphi_A)$. This action has a second derivative of zero for eigenstates of $\hat{A}$, indicating that even a fully projective measurement could be identified as ``weak'' in the sense that the back-action will not have any observable effect on the amplitude of the post-selected output. We can therefore conclude that the sensitivity of the post-selected state to the back-action dynamics generated by $\hat{A}$ is the critical factor that determines whether the disturbance caused by the measurement interaction can be neglected or not. 

To summarize the results of this section, we have derived a trade-off relation between the resolution error $\delta A_M$ and the uncertainty $\delta A_S$ of the value of $\hat{A}$ determined from initial and final conditions of the system caused by the back-action uncertainty of the measurement. The fluctuations of the meter readout associated with a specific set of initial and final conditions is minimized at $(\delta A_M)^2 = (\delta A_S)^2$, where the correlation between the weak values determined by $\ket{\psi}_S$ and $\bra{f}$ and the observable meter shift is optimal. For $(\delta A_M)^2 > (\delta A_S)^2$, the weak value provides an accurate description of the meter shift, but the high resolution error caused by meter noise makes it difficult to identify this correlation in the actual readout data. For $(\delta A_M)^2 < (\delta A_S)^2$, the meter readout provides more precise information about the value of $\hat{A}$ than the weak values obtained from the various back-action parameters $\varphi_A$. However, the meter shift is still described by the actions $S(\psi, f, \varphi_A)$ associated with every eigenstate $\ket{b}$ of the meter. The relation between meter shift and back-action dynamics thus changes from the Hamilton-Jacobi relation of Eq.(\ref{eq:HamiltonJacobi}) into a more complicated quantum superposition of very different back-action parameters $\varphi_A$. In the following, we will therefore consider how the value of $\hat{A}$ observed in the meter readout is determined by quantum interferences between the different back-action components in the dynamics of the system.

\section{Emergence of eigenvalues from quantum interferences between different back-action dynamics}
\label{sec:eigenvalues}
At high measurement resolutions, the information  obtained in a meter readout may be sufficient to determine the value of $\hat{A}$. The additional information provided by the final conditions $\bra{f}$ will play a much less significant role. It is therefore important to investigate how the role of the back-action dynamics changes as the measurement resolution increases. It should be noted that Eq.(\ref{eq:effect on M}) applies regardless of measurement strength. The effect of the system on the meter is completely defined by the combination of initial and final conditions, where the dependence of the amplitude of $\bra{f}$ on the back-action parameter $\varphi_A = g B_b$ defines the operator acting on the meter according to Eq.(\ref{eq:effect on M}). For sufficiently small uncertainties $\delta \varphi_A$ of the back-action parameter, the meter shift is fully determined by the $\varphi_A$-dependence of the phase of $\bra{f}$ given by the Hamilton-Jacobi equation Eq.(\ref{eq:HamiltonJacobi}). This is the reason why we were able to avoid a discussion of meter preparation and readout in the previous sections. In the limit of high measurement resolution, the initial meter state $\ket{\phi}_M$ will define the initial statistics of the back-action parameters $\varphi_A = g B_b$, and the meter readout will modify this distribution while also defining the information about $\hat{A}$ obtained from the readout result $\bra{m}$. Both characteristics of the measurement must be designed carefully to ensure that a valid measurement can be performed. For a meter initialized in a state $\ket{\phi}_M$ and read out in $\bra{m}$, the total probability amplitude is
\begin{equation}
\label{eq:amplitude}
\bra{f, m} \hat{U}_{SM} \ket{\psi_S, \phi_M} = \sum_b \bra{f} \hat{U}_{\mathrm{back}}(g B_b)\ket{\psi}_S \left(\braket{m|b} \braket{b|\phi}_M \right) .
\end{equation}
The probability of a meter readout $m$ is determined by quantum interferences between the amplitudes of the final condition $\bra{f}$ for different back-action parameters $\varphi_A = g B_b$. The distribution of measurement outcomes that emerges from these quantum interference effects will be very sensitive to the phases of the amplitudes of $\bra{f}$ as described by the action $S(\psi, f, \varphi_A)$. The range of possible back-action parameters $\varphi_A = g B_b$ is determined by the absolute values of the meter amplitudes. The square of these absolute values corresponds to the probability of finding first $b$ and then $m$ in a sequential projective measurement of the initial meter state $\ket{\phi}_M$,
\begin{equation}
\label{eq:meter}
P(b,m|\phi_M) = |\braket{m|b}|^2 |\braket{b|\phi}_M|^2.
\end{equation}
The initial meter state thus determines the statistics of the back-action. However, the actual purpose of the measurement is the evaluation of $\hat{A}$ from a meaningful meter readout $\bra{m}$. For this purpose, the meter readout should be maximally sensitive to the meter shift defined by a unitary transformation generated by $\hat{B}$. Each meter readout $\bra{m}$ can then be identified with a measurement outcome $A_m$ for the physical property $\hat{A}$ that is defined by the linear dependence of the phases of $\braket{m|b}$ on the eigenvalues $B_b$,  
\begin{equation}
\label{eq:meter shift}
\hbar \mbox{Arg}(\braket{m|b} \braket{b|\phi}_M)= g B_b \; A_m. 
\end{equation}
This relation describes an optimal meter readout and can be used to determine the statistics of a measurement with high resolution. In general, the distribution of the back-action parameters $\varphi_A=g B_b$ will depend on the precise form of the initial meter state and its readout as given by Eq.(\ref{eq:meter}). In order to achieve the ideal limit of a maximally resolved measurement, the uncertainties of $\hat{B}$ in the initial meter state $\ket{\phi}_M$ and in the readouts $\bra{m}$ should approach infinity. It is therefore possible to represent this ideal limit by a homogeneous distribution of all values of $\varphi_A$. Since the density of meter states must be sufficiently high in this limit, it is also possible to replace the sum over $b$ with an integral over $\varphi_A$,
\begin{equation}
\sum_b \left(\braket{m|b} \braket{b|\phi}_M \right) \to \int \frac{1}{\sqrt{2 \pi \hbar}} \exp\left(\frac{i}{\hbar} A_m \varphi_A \right) d\varphi_A.
\end{equation}
In this limit, Eq.(\ref{eq:amplitude}) corresponds to a Fourier transform of the $\varphi_A$-dependence of the amplitude of $\bra{f}$ for the input state $\ket{\psi}_S$,
\begin{equation}
\label{eq:Fourier transform}
\bra{f, m} \hat{U}_{SM} \ket{\psi_S, \phi_M} = \frac{1}{\sqrt{2 \pi \hbar}} \int 
\bra{f} \hat{U}_{\mathrm{back}}(\varphi_A) \ket{\psi}_S
\exp \left(\frac{i}{\hbar} A_m \varphi_A \right) d \varphi_A.
\end{equation}
The meter readout is therefore sensitive to the Fourier components of the back-action dynamics described by $\hat{U}_{\mathrm{back}}(\varphi_A)$. These Fourier components are given by projectors on the eigenstates $\ket{a} \bra{a}$,
\begin{equation}
\frac{1}{\sqrt{2 \pi \hbar}} \int 
\hat{U}_{\mathrm{back}}(\varphi_A) 
\exp \left(\frac{i}{\hbar} A_m \varphi_A \right) d \varphi_A = \sum_a \ket{a}\bra{a}\delta(A_m - A_a).
\end{equation}
For the distribution of meter readouts $m$, we obtain
\begin{equation}
\label{eq:eigenvalue}
\bra{f, m} \hat{U}_{SM} \ket{\psi_S, \phi_M} = \sum_a \braket{f|a} \braket{a|\psi}_S \delta(A_m - A_a).
\end{equation}
The delta function indicates that the only readout results $m$ with a non-vanishing probability are the ones where the value of the physical property of the system $A_m$ corresponds to an eigenvalue $A_a$ of $\hat{A}$. The effect of the measurement of $m$ on the system is given by a projection operator,
\begin{equation}
\label{eq:projector}
\bra{m} \hat{U}_{SM} \ket{\phi}_M = \sum_a  \delta(A_m - A_a) \ket{a} \bra{a}.
\end{equation}
Both Born's rule and the assumption that projective measurements leave the system in an eigenstate of the target observable after the measurement can thus be explained as an effect of the interferences between different back-action dynamics associated with the quantum coherence of meter preparation and readout. 

It seems important that the projective measurement only emerges as a limiting case of the more general dynamics of measurement interactions. Eigenvalues are determined by the periodicities of the dynamics generated by the physical property $\hat{A}$. It is therefore problematic to identify eigenvalues with the static physical properties of an undisturbed system. According to the quantum formalism describing the measurement interaction, eigenvalues appear in the meter readout because of quantum interferences between the unitary operations representing the change of the system associated with different back-action parameters $\varphi_A$. The quantum formalism thus identifies eigenvalues as elements of the back-action dynamics. All measurement outcomes can then be explained by the quantum interferences between different back-action dynamics, without any need for measurement independent elements of reality.

\section{Back-action effects in the Stern-Gerlach experiment}
\label{sec:SG}
It may be difficult to picture the physics of back-action effects described by the rather general formalism developed above. It may therefore be useful to consider the precise role of the back-action dynamics in the case of a Stern-Gerlach experiment, where the $z$-component of the spin of a spin-1/2 particle is measured by evaluating the momentum transfer to the particle as it passes through an inhomogeneous magnetic field. The target observable $\hat{A}$ is then given by the spin component $\hat{\sigma_z}$ with eigenvalues of $\pm\hbar/2$ for the eigenstates $\ket{\uparrow}$ and $\ket{\downarrow}$, and the generator of the meter response is given by the position coordinate $\hat{z}$ along the inhomogeneity of the magnetic field. The back-action can then be visualized as a spin precession around the $z$-axis, where the angle of the spin rotation is given directly by the back-action parameter $\varphi_A=g z$. The quantum amplitudes of the meter response can be expressed in terms of the corresponding unitary transformation of the spin, as defined in Eq.(\ref{eq:backaction}). For an initial state $\ket{\psi}_S$ that is an equal superposition of the spin-up and the spin-down state, these amplitudes are given by
\begin{equation}
\label{eq:SGamp}
\bra{f} \hat{U}_{\mathrm{back}}(\varphi_A) \ket{\psi}_S = \frac{1}{\sqrt{2}} \braket{f|\uparrow} \exp \left(-i \frac{\varphi_A}{2} \right) +  \frac{1}{\sqrt{2}} \braket{f|\downarrow} \exp \left(i \frac{\varphi_A}{2} \right).
\end{equation} 
This amplitude will modify the input state $\ket{\phi}_M$ of the meter, depending on the position $z$ at which the particle moves through the magnetic field. Typically, the meter state can be described by a Gaussian wavefunction, achieving a minimal momentum uncertainty through quantum coherence between the different positions $z$,
\begin{equation}
\label{eq:SGgauss}
\ket{\phi}_M = \int_{-\infty}^{\infty} \left(\frac{2 \Delta p_z^2}{\pi \hbar^2}\right) \exp\left(-\frac{\Delta p_z^2}{\hbar^2} z^2\right) \ket{z} dz.
\end{equation}
The spin precession at each location $z$ modifies the probabilities of subsequent measurements $\bra{f}$ of the spin. Here, we consider subsequent measurement outcomes $\bra{f}$ that are maximally sensitive to the initial spin orientation of $\ket{\psi}_S$. This means that their spin directions will lie in the $xz$-plane, as represented by real values amplitudes $\braket{f|\uparrow}$ and $\braket{f|\downarrow}$. The conditional probability of finding the outcome $\bra{f}$ after the particle passed through the magnetic field at $z=\varphi_A/g$ is then given by
\begin{equation}
P(\psi, f, \varphi_A) = \frac{1}{2}(1+ 2 \braket{f|\uparrow}\braket{f|\downarrow} \cos(\varphi_A)).
\end{equation}
If both coefficients are positive, the probability is maximal for $\varphi_A=0$ and minimal for $\varphi_A=\pi$. Effectively, the subsequent measurement provides us with information about the meter observable $\hat{z}$ by identifying the magnitude of the magnetic field through which the particle passed. On the other hand, the meter response is encoded in the quantum phases associated with the action of the spin dynamics,
\begin{equation}
S(\psi, f, \varphi_A) = - \hbar \arctan\left(\frac{|\braket{f|\uparrow}|^2-|\braket{f|\downarrow}|^2}{1+ 2 \braket{f|\uparrow}\braket{f|\downarrow}} \tan\left(\frac{\varphi_A}{2}\right)\right).
\end{equation}
It is important to note that this action defines $z$-dependent meter shifts, even if the meter was prepared with a very large $z$-uncertainty. Even after the interaction, it is possible to read out the value of $z$ with arbitrary precision. In a joint readout of $z$ and $p_z$ using an uncertainty limited quantum measurement with Gaussian resolutions, it is possible to satisfy the conditions of the weak measurement limit by including the back-action associated with the $z$-readout in the evaluation of any subsequent measurement outcomes $\bra{f}$. As discussed in Sec. \ref{sec:metershift}, the back-action dependent meter shift is given by the Hamilton-Jacobi equation associated with the back-action dynamics. According to Eq.(\ref{eq:HamiltonJacobi}), the value of the spin component $\hat{\sigma}_z$ associated with the average change of the momentum $p_z$ is given by
\begin{equation}
- \frac{\partial S(\psi,f,\varphi_A)}{\partial \varphi_A} = \frac{\hbar}{2} \frac{W_0}{(\cos(\varphi_A/2))^2+ W_0^2 (\sin(\varphi_A/2))^2},
\end{equation}
where
\begin{equation}
W_0=\frac{\braket{f|\uparrow}-\braket{f|\downarrow}}{\braket{f|\uparrow}+\braket{f|\downarrow}}
\end{equation}
corresponds to the weak value of $\hat{\sigma}_z$ at $\varphi_A=0$ in units of $\hbar/2$. The reason for the dependence of the meter shift on the back-action parameter $\varphi_A$ can be found in the modification of the relation between the initial state $\ket{\psi}_S$ and the subsequent measurement outcome $\bra{f}$. At $\varphi_A=0$ and at $\varphi_A=\pi$, extremal values are obtained because $\bra{f}$ determines a specific relation between $\hat{\sigma}_z$ and the $x$-component of the spin defined by the initial state $\ket{\psi}_S$. A spin rotation by $\pi$ inverts this relation, resulting in opposite extremes of the possible meter shifts. For other back-action angles $\varphi_A$, the relation between $\bra{f}$ and $\hat{\sigma}_z$ is not as strong.

As shown in the previous section, the eigenstates of $\hat{\sigma}_z$ only appear in the meter readout when that readout corresponds to a Fourier transform of the back-action components. In the present case, this would be a readout of the momentum $p_z$. As indicated in Eq.(\ref{eq:SGgauss}), the precision of the readout is then only limited by the momentum uncertainty $\Delta p_z^2$ in the initial meter state $\ket{\phi}_M$. If the values of $p_z$ associated with the eigenvalues of $\pm \hbar/2$ are much larger than $\Delta p_z^2$, the readout distribution around each eigenvalue result can be identified with the delta functions given in Eqs.(\ref{eq:eigenvalue}) and (\ref{eq:projector}). Oppositely, the delta functions in these equations always represent the sufficiently narrow distributions of the readout observable in the initial state. Importantly, these narrow readout distributions are only obtained when the readout measurement on the meter defines a quantum interference of all possible back-action parameters $\varphi_A$, effectively erasing all available information on the specific back-action experienced by the system. In the case of a Stern-Gerlach experiment, this is information about the precession angle of the spin. The loss of this information limits the information on $\hat{\sigma}_z$ that can be obtained in subsequent measurements of the spin. The reduced sensitivity of the measurement statistics to post-selection should therefore be traced to the loss of back-action information caused by a specific readout of the meter.

\section{Back-action effects in the final condition}
\label{sec:finalcondition}
The purpose of a measurement is to obtain information about the target observable $\hat{A}$. However, the quantum formalism make it difficult to separate the information about $\hat{A}$ from the back-action dynamics of the system. Quantum states only provide reliable information about the true value of $\hat{A}$ when they are eigenstates of the operator $\hat{A}$ and are therefore not affected by the back-action dynamics. The general case is described by initial conditions $\ket{\psi}_S$ and final conditions $\bra{f}$, neither of which is given by an eigenstate of $\hat{A}$. We can then show that the meter shift is determined by the effects of the back-action dynamics on the relation between  the initial condition $\ket{\psi}_S$ and the final condition $\bra{f}$. This relation has a classical analog, as shown by the Hamilton-Jacobi equation (\ref{eq:HamiltonJacobi}). The reason why $A(\psi, f, \varphi_A)$ depends on the back-action parameter $\varphi_A$ is that the back-action has an effect on all future outcomes $\bra{f}$. Therefore, the same outcome $\bra{f}$ determines a different value of $\hat{A}$ for different parameters $\varphi_A$.

If the precise value of $\varphi_A$ is known, the final condition $\bra{f}$ provides the necessary information for an independent evaluation of $\hat{A}$. In the quantum mechanical limit, the result of the Hamilton-Jacobi equation (\ref{eq:HamiltonJacobi}) is the weak value of $\hat{A}$ determined by the initial state $\ket{\psi}_S$, the dynamics $\hat{U}_{\mathrm{back}}(\varphi_A)$ and the final state $\bra{f}$. The weak value is thus obtained as the best estimate of the value of $\hat{A}$ under the initial and final conditions selected for the analysis of the measurement \cite{Hal04}. However, the weak value also depends on the back-action parameter $\varphi_A$ and this parameter can only be known with precision when the meter readout does not resolve the value of $\hat{A}$ at all. A meter readout with a low resolution error $\delta A_M$ is only possible when the back-action parameter $\varphi_A$ is unknown. In this regime, it is impossible to determine the value of $\hat{A}$ from the relation between the initial condition $\ket{\psi}_S$ and the final condition $\bra{f}$, because the unknown back-action dynamics given by $\varphi_A = g B_b$ is a necessary part of this relation. The back-action uncertainty thus erases part of the information that would be necessary to determine the value of $\hat{A}$ form a final measurement of $\bra{f}$.

It may be important to note that the general formalism describes all meter shifts as coherent superpositions of meter state components $\ket{b}$ as shown by the general form of Eq.(\ref{eq:effect on M}). The meter readout is always described by the interferences between different back-action dynamics described by coherent amplitudes of the final conditions $\bra{f}$ for different values of $\varphi_A=g B_b$. Weak values are obtained when the meter shift is described by quantum interferences in a narrow range of parameter values $\varphi_A$ so that the main difference between the amplitudes associates with different values of $\varphi_A$ can be described by the Hamilton-Jacobi equation (\ref{eq:HamiltonJacobi}). The only difference between the weak value limit and larger uncertainties in the back-action is that the interference effects get harder to evaluate and to interpret. 

To correctly understand the relation between the meter readout and the coherences in the back-action dynamics, it is necessary to remember that the interaction entangles the meter and the system. This means that the final measurement of the system $\bra{f}$ determines the coherences of the meter state and therefore the possible values of the meter shift. This dependence of the meter shift on the final condition $\bra{f}$ even applies in strong interactions that maximally entangle the system and the meter. If the final condition $\bra{f}$ is determined before the meter readout, the meter state is described by
\begin{equation}
\label{eq:meterstate}
\bra{f} \hat{U}_{SM} \ket{\psi_S, \phi_M} = \sum_b \bra{f} \hat{U}_{\mathrm{back}}(b B_b)\ket{\psi}_S  \braket{b|\phi}_M \ket{b}.
\end{equation}
It is hard to see any trace of an eigenvalue in this meter state. If it is read in the basis $\{ \ket{b} \}$, it is still possible to determine the back-action $\varphi_A$. If the coherence between only two neighboring values of $B_b$ are probed, it may be possible to observe the weak value of $A(\psi, f, \varphi_A)$. Eigenvalues only appear in the meter readout when the readout is described by a coherent superposition of a sufficiently wide range of eigenstates $\ket{b}$. As explained in Sec. \ref{sec:eigenvalues}, the eigenvalues of $\hat{A}$ are then obtained from the Fourier components of the $\varphi_A$-dependence of the amplitude of $\bra{f}$ in Eq.(\ref{eq:meterstate}).

Even in the limit of strong interactions, it is not possible to assign a readout independent value to $\hat{A}$. Instead, the future condition $\bra{f}$ of the system defines conditional states that can still be read out in many ways. Our analysis indicates that the conditions under which eigenvalues are observed are rather extreme and highly restrictive. The observation of eigenvalues requires a particularly strong coherence between different back-action parameters that can only be obtained when both the initial meter state and its readout support this coherence. The quantum formalism suggests that different selections of interactions and readout can change the values of the target property $\hat{A}$ from weak values associated with specific back-action parameters $\varphi_A$ into the eigenvalues determined by the Fourier components of the back-action dynamics given by $\hat{U}_{\mathrm{back}}(\varphi_A)$. We believe that this change in the character of measurement outcomes corresponds to the definition of a measurement context by a combination of interaction and readout. For weak interactions, the measurement context is determined by the final condition $\bra{f}$. For strong interactions, the context is determined by the meter readout. A context independent explanation of measurement outcomes is impossible because the interaction necessarily entangles the possible readouts with the future measurement contexts given by $\bra{f}$. In this sense, system-meter entanglement should be understood as a representation of contextuality in quantum measurements \cite{Auf19}.

\section{Conclusions}
\label{sec:conclude}
The representation of quantum measurements by Born's rule is problematic because it suggests that the eigenstate components of a quantum state represents an intrinsic measurement independent reality. In fact, this kind of reality only emerges after an interaction of the system with a sufficiently sensitive environment \cite{Mat21}. In the present paper, we have analyzed the way in which the dynamics of the system determines the precise effects of an interaction on a meter system. The response of the meter can be described entirely in terms of the parametrized evolution of the system caused by the back-action, where a final condition $\bra{f}$ is needed for maximal information on the transformation of the meter system. This final condition is a summary of all information that might be obtained after the measurement interaction has been completed. Our results show that this information changes the way in which the system acts on the meter by defining specific quantum coherences between the eigenstates of $\hat{A}$. We find that the meter shift is precisely defined by the weak value $A(\psi, f, \varphi_A)$ for initial condition $\ket{\psi}_S$, the final condition $\bra{f}$ and the back-action $\varphi_A$. Weak values provide an accurate description of the meter shift if the back-action uncertainty is sufficiently small. The reason why weak values cannot be observed in a single measurement is explained by the increases in the back-action uncertainty at high measurement resolutions. At high resolution, the meter readout selects specific quantum interferences between the different back-action dynamics. In the limit of large back-action uncertainties, these quantum interferences correspond to Fourier transforms of the back-action dynamics represented by the coherent amplitudes of the final condition $\bra{f}$. We thus find that the observation of an eigenvalue is the result of quantum interferences between a wide range of different back-action dynamics.

It may be important to note that there are alternative descriptions of quantum interference effects in the measurement process. It has been pointed out that weak values can be understood as quantum interference effects between different eigenstates of the target observable \cite{Dre15, Sok16}. In the present analysis, this interference between different eigenstates of $\hat{A}$ is included in the description of the back-action dynamics. Paraphrasing the title of \cite{Sok16}, it might be said that all quantum dynamics is quantum interference and very little else. However, this does not change the fact that the observable information about the dynamics can only be obtained in a subsequent measurement that defines the coherences between the different eigenstates of $\hat{A}$ in terms of a single observable outcome $\bra{f}$. The quantum interference effects between different eigenstates of $\hat{A}$ are therefore summarized by the back-action dynamics.

Our analysis shows that all measurement results originate from quantum interferences between different magnitudes of the back-action parameter $\varphi_A$ associated with the meter observable $\hat{B}$. Although it is technically correct that anomalous weak values are caused by disturbances of the system \cite{Ips22}, the effect of the back-action dynamics is minimal in the case of weak interactions. Weak values represent quantum interferences of a very small range of different values of the back-action parameter $\varphi_A$. The effect of quantum coherences between different back-action dynamics is much stronger in the case of eigenvalues. The analysis presented in this paper shows that eigenvalues can be understood as characteristic signatures of the back-action dynamics that only emerge when the disturbance covers a sufficiently wide range of different values of the back-action parameter $\varphi_A$. In the light of this result, it seems necessary to conclude that eigenvalues are only observed as a result of the disturbance of the system in a quantum measurement and cannot be identified with measurement independent realities. The quantum formalism does not provide any measurement independent definition of physical properties. Instead, the relation between different measurements is determined entirely by the dynamics of the system. A proper understanding of quantum mechanics is only possible if this dynamical relation between different measurement contexts is taken into account.

\section*{acknowledgment}
This work was supported by JST, the establishment of university fellowships towards the creation of science technology innovation, Grant Number JPMJFS2129.

\end{document}